\documentclass[12pt,nofootinbib]{article}
\usepackage[dvips]{graphicx}
\usepackage{amssymb}

\usepackage{amsmath}
\usepackage{epsfig}

\numberwithin{equation}{section}

\def\beq{\begin{equation}}
\def\eeq{\end{equation}}
\def\be{\begin{equation}}
\def\ee{\end{equation}}
\def\bea{\begin{eqnarray}}
\def\eea{\end{eqnarray}}
\def\d{{\rm d}}

\def\be{\begin{equation}}
\def\ee{\end{equation}}
\def\beq{\begin{equation}}
\def\eeq{\end{equation}}
\def\bea{\begin{eqnarray}}
\def\eea{\end{eqnarray}}

\newcommand{\dd}{{\rm d}}
\newcommand{\Muv}{M_{\rm UV}}

\DeclareRobustCommand{\SkipTocEntry}[4]{}

\begin{document}

\begin{titlepage}

\setcounter{page}{1} \baselineskip=15.5pt \thispagestyle{empty}

\begin{flushright}
{\footnotesize ITEP-TH-17/08}\\
{\footnotesize SU-ITP-08/19}\\
{\footnotesize SLAC-PUB-13365}\\
{\footnotesize PUPT-2277}
\end{flushright}
\vfil

\bigskip\
\begin{center}
{\LARGE  Holographic Systematics of D-brane Inflation}
\vskip 15pt
\end{center}

\vspace{0.5cm}
\begin{center}
{\large Daniel Baumann,$^{1,2}$ Anatoly Dymarsky,$^{3}$ Shamit Kachru,$^{3,4}$ }
\vskip 3pt
{\large Igor R. Klebanov,$^{2,5}$ and Liam McAllister$^{6}$}
\end{center}

\vspace{0.3cm}

\begin{center}
\textit{${}^1$ Department of Physics, Harvard University,
Cambridge, MA 02138}\\

\vskip 4pt
\textit{${^2}$ Department of Physics, Princeton University,
Princeton, NJ 08544}\\

\vskip 4pt
\textit{${}^3$ Department of Physics, Stanford University,
Stanford, CA 94305}\\

\vskip 4pt
\textit{${}^4$ SLAC, Stanford University,
Stanford, CA 94309}\\

\vskip 4pt
\textit{${}^5$
Center for Theoretical Science, Princeton University,
Princeton, NJ 08544}\\

\vskip 4pt
\textit{${}^6$ Department of Physics, Cornell University,
Ithaca, NY 14853}
\end{center} \vfil

\vspace{0.8cm}

\noindent
We provide a systematic treatment of possible corrections to the inflaton potential for D-brane inflation in the warped deformed conifold. We consider the D3-brane potential in the presence of the most general possible corrections to the throat geometry sourced by coupling to the bulk of  a compact Calabi-Yau space.  This corresponds to the potential on the Coulomb branch of the dual gauge theory, in the presence of arbitrary perturbations of the Lagrangian.  The leading contributions arise from perturbations by the most relevant operators that do not destroy the throat geometry.
We find a generic contribution from a non-chiral operator of dimension $\Delta=2$ associated with a global symmetry current, resulting in a negative contribution to the inflaton mass-squared.  If the Calabi-Yau preserves certain discrete symmetries, this is the dominant correction to the inflaton potential, and fine-tuning of the inflaton mass is possible.  In the absence of such discrete symmetries, the dominant contribution comes from a chiral operator with $\Delta=3/2$, corresponding to a $\phi^{3/2}$ term in the inflaton potential.
The resulting inflationary models are phenomenologically
%identical
similar
%DB
to the inflection point scenarios arising from specific D7-brane embeddings, but occur under far more general circumstances.
Our strategy extends immediately
to other warped geometries, given sufficient knowledge of the Kaluza-Klein spectrum.

\vfil

\end{titlepage}

\newpage
\tableofcontents

\newpage
\section{Introduction}

\vskip 5pt
\subsection{Motivation}
\vskip 4pt
Important conceptual problems of the standard Big Bang cosmology, including the horizon and flatness problems, are resolved if the very early universe underwent a brief phase of accelerated expansion \cite{Guth}.  In addition to explaining the large-scale homogeneity of the universe, such an inflationary epoch provides a quantum-mechanical mechanism to generate primordial inhomogeneities, which are required for the formation of structure and have been observed as anisotropies in the temperature of the cosmic microwave background (CMB).
A statistical analysis of the primordial fluctuations inferred from the recent CMB data \cite{Komatsu} is in good agreement with the basic expectations from inflation \cite{Guth, Linde, Steinhardt}.

Given the phenomenological success of inflation,
an important current direction in theoretical physics seeks to
reveal the microphysical origin of the accelerated expansion.
Since effective field theory models of inflation are sensitive to assumptions about the ultraviolet (UV) structure of the theory, it is instructive to
build controlled and predictive inflationary models in the general framework of string theory \cite{LiamEva}.

One well-studied class of string inflation models involves motion of a D3-brane towards an anti-D3-brane
in a warped throat region of a (conformally)
Calabi-Yau flux compactification \cite{KKLMMT}. An explicit example of such a geometry is the warped deformed conifold \cite{KS,KT}.   The general form of the potential for the inflaton
$\phi$, which is the scalar describing the radial position of the D3-brane in the throat, is
\begin{equation}
\label{basic}
V (\phi)~=~V_{D3/\overline{D3}}(\phi) ~+~ H^2 \phi^2 ~+~ \Delta V(\phi)\, ,
\end{equation}
where $V_{D3/\overline{D3}}(\phi)$ includes the warped antibrane tension and the Coulomb potential attracting the D3-brane to the anti-D3-brane, and the mass term, $H^2 \phi^2$, arises from effects related to moduli stabilization.  These terms
were computed in \cite{KKLMMT}.
The large mass of order the Hubble parameter, $H^2 \approx V/(3 M_{\rm pl}^2)$, generically spoils inflation, unless it can be cancelled against other effects.

The third term in the potential, $\Delta V$, represents all possible additional effects.  In particular, $\Delta V$ includes  the corrections to the potential that arise from embedding the system into a compact Calabi-Yau space: one expects bulk fluxes, distant branes, further moduli-stabilization effects, and so forth to perturb the throat geometry, resulting
in some change to the inflaton potential.
It has been natural to think that
by including the effects captured by $\Delta V$, one will sometimes find small-field models of D-brane
inflation by fine-tuning.  However, it has remained challenging to encompass all such contributions to the potential within a single computable framework.\footnote{For example, in \cite{delicate,KP} the effects of D7-branes in the throat region were incorporated, but the effects of the bulk Calabi-Yau space were not.}
In this paper, we present an effective parametrization
of $\Delta V$, {\it i.e.} of
%all possible
the leading
%DB
compactification effects in warped D-brane inflation.

The compactified throat geometry can be
described, via AdS/CFT duality, as an (approximately) conformal
field theory (CFT) which is cut off at some high mass scale $M_{\rm UV}$ and coupled to the Calabi-Yau moduli
and to four-dimensional supergravity.   Our prescription involves perturbing the CFT Lagrangian by the leading irrelevant operators
that
generate a Coulomb-branch potential for the CFT fields.
Since the D3-brane position $\phi$ is precisely such a Coulomb branch vev in the CFT, this gives an effective method for computing $\Delta V$.
Moreover, the dual gravity description allows for an efficient
determination of the full list of such perturbations, through study of the spectrum of Kaluza-Klein (KK) modes in the
gravitational background.
For excitations around the $AdS_5\times T^{1,1}$ background, the KK modes have been
classified in \cite{Gubser,Ferrara} and matched to gauge-invariant operators in the dual ${\cal N}=1$ superconformal
field theory \cite{KW,Morrison}.
Armed with these results,
we determine the structure of D-brane inflation in the specific case of the conifold throat.
Our method generalizes immediately to
inflationary models in any geometry that is locally approximated by a Calabi-Yau cone.

\subsection{The Inflaton Potential from Gauge/Gravity Duality}

Let us now spell out our strategy in greater detail.  For the local throat geometries arising in gauge/string duality, such as the warped deformed conifold \cite{KS}, there is a very natural way to list the leading corrections to the inflaton potential. When one compactifies the throat, one should expect to
perturb the throat geometry in the UV region by the leading irrelevant operators present
in the coupling of the dual CFT to the Calabi-Yau moduli sector.\footnote{Relevant perturbations of the throat
are of course possible in general; we have excluded them by the assumption that a local region
of the flux compactification is well-modeled by the noncompact throat.  This seems to be sensible
in full string compactifications, and is
the same assumption that is always made in this
class of models.}  Among these operators, some induce a potential on the Coulomb branch, and thus contribute to the inflaton potential.  The corresponding statement in the supergravity language is that we are interested in the leading UV deformations of the background that
affect the potential of a probe D3-brane; perturbations of fields to which the D3-brane does not couple directly may be ignored in our linearized treatment.

 To proceed further, it is helpful to make some concrete assumptions about the global features of the
 string compactification; we emphasize, however, that our strategy could be generalized to a wide variety of other scenarios.  We assume the existence of a warped throat region glued into a compact Calabi-Yau space.  Throughout this paper, we suppose that the throat is long, so that the gauge theory is approximately conformal across a wide range of energy scales.  Moreover, we consider a D3-brane that is well-separated both from the UV and from the infrared (IR) regions. Much of our analysis will refer to the specific example of the warped deformed conifold throat \cite{KS},
 which asymptotically approaches $AdS_5\times T^{1,1}$, up to factors logarithmic in the AdS radius \cite{KT}. Since these factors vary slowly, an approximation that is quite useful is to study D3-brane motion in the exactly conformal background $AdS_5\times T^{1,1}$.

 We assume that the moduli are stabilized as in \cite{KKLT} or its variants.
 Supersymmetry is broken by the presence of the anti-D3-brane at the end of the throat \cite{Kachru:2002gs};
 if the minimal warp factor in the throat is $a_0$, this sources a vacuum energy $\sim 2 a_0^4 T_3$, where
 $T_3$ is the brane tension.

 In this kind of scenario, we argue in Appendix A that quite generally, there are bulk moduli fields
%\footnote{Strictly speaking, $X$ could instead be an open string modulus superfield whose lowest component is the inflaton $\phi$ itself, as we explain in Appendix A.}
 $X$ with F-terms\footnote{The scale of $F_X$ is constrained by the requirement that the energy associated with the UV perturbation must not lead to decompactification of the compact space (see Appendix A).} $F_X \sim \xi \, a_0^2$.
 %DB new footnote
We implicitly assume that the value of the coefficient $\xi$ can be fine-tuned by considering different bulk fluxes, different bulk sources of SUSY-breaking, or even distinct Calabi-Yau geometries. (This is of course the standard procedure suggested by Wilsonian effective field theory.)
%this is very plausible due to the non-renormalization theorem.  Then 
The coupling of the CFT fields to the bulk moduli will result in a leading perturbation to the K\"ahler potential\footnote {We postpone to a future publication a complete treatment of direct superpotential couplings between the CFT fields and $X$, as their effects can be somewhat subtle \cite{Gminus}.} ${\cal K}$, and correspondingly to the scalar potential $V$, of the form:
  \be
 \label{mainpoint}
 \Delta {\cal K} =  c \int d^4 \theta \
 M_{\rm UV}^{-\Delta}\ X^\dagger X
 \, {\cal O}_{\Delta}\qquad \Rightarrow \qquad
 \Delta V =
 c~M_{\rm UV}^{-\Delta}
 |F_X|^2\, {\cal O}_{\Delta}\, ,
 \ee
 where
 ${\cal O}_{\Delta}$ is a gauge-invariant operator of dimension $\Delta$
 in the (approximately) conformal gauge theory dual to the throat, and $c$ is a constant.
 The scale $\Muv$ relates to the UV cutoff of the gauge theory or equivalently to the large $r$ limit of the throat geometry.

 As noted above, the operators ${\cal O}_\Delta$ of interest are built out of scalar fields, so that they create a potential on the Coulomb branch of the gauge theory.  In  particular, we will be interested in contributions to the potential for radial motion, with corresponding scalar field $\phi$. We will show that for suitable ${\cal O}_\Delta$, (\ref{mainpoint}) induces a radial potential of the form\footnote{For simplicity of presentation we restrict these preliminary remarks to perturbations induced by a single operator. Perturbations to the Lagrangian induced by more than one operator will be presented below.}
 \beq \label{equ:simplepoint}
   \Delta V =
 -c~\Muv^{-\Delta}
 |F_X|^2\, {\phi}^{\Delta}  \, ,
\eeq
with $c>0$.  The overall minus sign in (\ref{equ:simplepoint}) arises after a minimization in the angular directions. This potential contributes an expulsive force driving the D3-brane towards the UV.

To parametrize the leading corrections of this sort to the D3-brane potential, one is therefore interested in the handful of lowest-dimension operators that produce a potential on the Coulomb branch.
Happily, the supergravity modes around $AdS_5\times T^{1,1}$ have been classified \cite{Gubser,Ferrara} and matched to the dimensions
 of the corresponding operators in the `conifold CFT' \cite{KW,Morrison}, {\it i.e.}~the $SU(N)\times SU(N)$ gauge theory coupled to bi-fundamental fields
 $A_i, B_j$ ($i,j=1,2$).
We will use these results to analyze the D3-brane potential from the gravity and gauge theory points of view in \S2 and \S3, respectively.

We will see there that the lowest-dimension contributing operators ${\cal O}_\Delta$ are chiral operators of dimension 3/2 and non-chiral operators of dimension 2; the latter are
protected from acquiring large anomalous dimensions as they are in supermultiplets of global symmetry currents.\footnote{\label{foot} When the throat is embedded into a compact Calabi-Yau, all continuous global symmetries are broken.
However, the resulting
 shift of the
operator dimension away from 2 is negligible in the relevant limit of a long throat.}
The chiral operators can be forbidden from appearing in terms of the form (\ref{mainpoint}) if
the compactification preserves a small discrete symmetry group, as explained further in \S4, but are otherwise generically
present.
Depending on the symmetries of the compactification, one then encounters two natural possibilities:

\vskip 4pt
 $\mathbf{1< \Delta< 2}$ {\bf (`fractional') case}:
 The perturbation (\ref{mainpoint}) involving the chiral operator
${\cal O}_{3/2}$ is present in the theory, {\it i.e.}~allowed
by all the symmetries.  Then, one will obtain a potential of the form
$\Delta V =- c\, M_{\rm UV}^{-3/2} |F_X|^2 \phi^{3/2}$.

\vskip 4pt
{\bf $\mathbf{\Delta=2}$ (quadratic) case}:
The operator ${\cal O}_{3/2}$ is forbidden from appearing in terms of the form (\ref{mainpoint})
by a discrete symmetry in the bulk, or
by orbifolding the conifold
as in {\it e.g.}~\cite{Morrison}.
Then  the leading perturbation comes from the non-chiral operator ${\cal O}_{2}$ and yields a potential of the form
$\Delta V =-c\, M_{\rm UV}^{-2} |F_X|^2 \phi^2$.

\vskip 4pt
The phenomenology, after appropriate fine-tuning to obtain inflation,
depends crucially on whether the fractional or quadratic case is realized; see \S5.
 The potential in the fractional case is of the same inflection-point form as the D3-brane potential generated by a moduli-stabilizing D7-brane stack that descends into the throat region while wrapping a suitable (Kuperstein-embedded \cite{Kuperstein}) four-cycle \cite{delicate, KP}.
However, from the present viewpoint, we see that this structure arises far more generally, and can be understood from the geometry of the throat itself, even in the absence of D7-branes in the throat region.
Therefore, the correction to the D3-brane potential induced by D7-branes as computed in \cite{delicate, KP}, is in fact
the most generic leading correction one would expect in a warped conifold throat coupled to an arbitrary
compact Calabi-Yau space.

In the quadratic case, which is also a fairly generic possibility, the leading correction $\Delta V$ induced by coupling to the bulk,
is an inflaton mass.  In this case, one should expect to find inflation only when one balances $\Delta V$ against the
$H^2\phi^2$ term in (\ref{basic}).  The resulting model will have phenomenology of the sort envisioned in
\cite{KKLMMT}, and analyzed in detail in \cite{FT}.

We should note that in other confining throat geometries, the harmonic analysis
of the KK spectrum would differ, and the corresponding specific form(s) of the potential, when it is suitable for inflation,
would likewise change.  However, our strategy would allow one to determine the structure of the D3-brane potential in a general throat, given sufficient information about the KK spectrum;
we briefly discuss other examples in  \S\ref{sec:general}.

\section{Potential for a D3-brane in a Warped Throat}

\subsection{D3-brane Potential in Type IIB Supergravity}

\noindent
{\it Warped flux compactification}
\vskip 3pt

We work in the supergravity approximation in which the Einstein-frame action for type IIB string theory takes the form
\begin{eqnarray}
\label{equ:action}
S_{\rm IIB} &=& \frac{1}{2 \kappa_{10}^2} \int \dd^{10} x \sqrt{|g|} \left[ {\cal R} + \frac{\partial_M \tau \partial^M \bar \tau}{2 \, {\rm Im}(\tau)^2} - \frac{G_3 \cdot \bar G_3}{12 \, {\rm Im}(\tau)} - \frac{\tilde F_5^2}{4 \cdot 5!} \right] \nonumber \\
&& + \frac{1}{8 i \kappa_{10}^2} \int \frac{C_4 \wedge G_3 \wedge \bar G_3}{{\rm Im}(\tau)} + S_{\rm local}\, ,
\end{eqnarray}
where $\kappa_{10}$ is the ten-dimensional gravitational coupling.
Here $\tau \equiv C_0 + i e^{- \phi}$ is the axio-dilaton field and $G_3 \equiv F_3 - \tau H_3$ is a combination of the RR and NS-NS three-form fluxes $F_3 \equiv \dd C_2$ and $H_3 \equiv \dd B_2 $.
The five-form $\tilde F_5 \equiv F_5 - \frac{1}{2} C_2 \wedge H_3 + \frac{1}{2} B_2 \wedge F_3$ is self-dual,
\beq
\tilde F_5 = \star_{10} \tilde F_5\, .
\eeq
$S_{\rm local}$ denotes contributions from D-branes and orientifold planes.
We will be interested in the following class of metrics:
\beq
\label{equ:warped}
\dd s^2 = e^{2A(y)} g_{\mu \nu} \dd x^\mu \dd x^\nu +
 e^{-2 A(y) } \tilde g_{mn}(y) \dd y^m \dd y^n\, ,
\eeq
where $y$ are the coordinates of the six-dimensional internal space.
In the special case of fluxes that are purely imaginary self-dual, $\star_6 G_3 = i G_3$, the metric $\tilde g_{mn}$
is simply the unwarped Calabi-Yau metric \cite{Giddings}. However,
we will be interested
in situations where more general fluxes, moduli-stabilization effects, and antibranes create more complicated warping of the internal manifold. Then $\tilde g_{mn}$ is not a Calabi-Yau metric; it has some additional warping and squashing \cite{DKM, Butti,DKS}.

The self-duality lets us write the five-form as
\beq
\tilde F_5 = (1+ \star_{10}) \left[ \dd \alpha(y) \wedge \, \dd x^0 \wedge \dd x^1 \wedge \dd x^2 \wedge \dd x^3 \right]\, .
\eeq
It will prove convenient to define the following combination of the warp factor and the five-form
\beq \label{equ:phim}
\Phi_\pm \equiv e^{4A} \pm \alpha\, ,
\eeq
and to decompose the three-form flux into imaginary self-dual (ISD) and imaginary anti-self-dual (IASD) components,
\beq
G_\pm \equiv (i \pm \star_6) G_3\, .
\eeq
The Einstein equations and the Bianchi identity for the five-form then imply\footnote{This corrects the corresponding equation in \cite{Giddings} by a numerical factor.}
\beq
\label{equ:PhiEoM}
 \widetilde \nabla^2 \Phi_\pm= \frac{e^{8A+ \phi}}{24}  |\widetilde{G_\pm}|^2 + e^{-4 A} |\widetilde \nabla \Phi_\pm|^2 + {\rm local} \, ,
\eeq
where $g_s \equiv e^{\phi} = 1/{\rm Im}(\tau)$ and tildes indicate that all contractions are with respect to the reference metric, $\tilde g_{mn}$.
In compactifications with ISD fluxes
($G_- = 0$),
$\tilde g_{mn}$ is a Calabi-Yau metric and
the tree-level flux solution is $\alpha = e^{4A}$, so that
$\Phi_- = 0$ \cite{Giddings}.
In this work we study linearized perturbations around this background solution.

In the perturbed solution, the internal space is squashed in such a way that the overall metric
 cannot be written as a
Ricci-flat metric with a single warp factor \cite{Butti}.
Nevertheless, the definition (\ref{equ:phim}) of $\Phi_-$ remains applicable: the $e^{4A}$ appearing in $\Phi_-$ is
the square of the warp factor multiplying the Minkowski metric, as indicated in (\ref{equ:warped}), while $\alpha=C_{0123}$, which is well-defined.
Moreover, at linear order $\Phi_-$ obeys the Laplace equation with respect to the unperturbed Calabi-Yau metric $\tilde g_{mn}^{(0)}$: perturbations in $\tilde g_{mn}$ correct the $\Phi_-$ equation of motion only at second order.
We defer a detailed study of the effects of $G_-$ perturbations to a separate publication~\cite{Gminus}.
Here, we study the solutions of (\ref{equ:PhiEoM}) without the $G_-$ source term, {\it i.e.}~the solutions of the Laplace equation
\beq
\tilde g^{mn}_{(0)}\nabla_m \nabla_n \Phi_-  \equiv \widetilde \nabla^2 \Phi_- = 0.
\eeq

\vskip 8pt
\noindent
{\it D3-brane potential and perturbations of the geometry}
\vskip 3pt

From the Dirac-Born-Infeld (DBI) and the Chern-Simons (CS) terms in the action for a D3-brane, one can see that the potential felt by a D3-brane
is
\begin{equation}
\label{phiminus}
V = T_{3} \, \Phi_{-} \, .
\end{equation}
To systematically investigate
the D3-brane potential, one is therefore interested in perturbations to the throat background that involve non-vanishing
values of the $\Phi_-$ field.  There are two different kinds of perturbations that might arise:

\begin{enumerate}
\item[a)]
perturbations of the throat by {\it normalizable} modes of the supergravity fields.
In the gauge theory language, these are perturbations of the
{\it state} of the dual CFT and correspond to SUSY-breaking perturbations in the IR that give rise to a D3-brane potential. Examples
include the SUSY-breaking antibrane state \cite{Kachru:2002gs} whose supergravity solution was found in  \cite{DKM}, or the baryonic branch state described in \cite{Butti,DKS}.
\item[b)]
perturbations of the throat by {\it non-normalizable}\footnote{In the full
compact solution these modes become normalizable, but their behavior in the limit of a noncompact throat is  that of non-normalizable modes.  Correspondingly, the modes referred to as normalizable in (a) are normalizable even in the noncompact limit.} modes of the supergravity fields.
These are deformations of the CFT {\it Lagrangian} induced by the coupling of the CFT to UV physics.
\end{enumerate}

The perturbations of type (a) due to the IR SUSY-breaking state have already been accounted for in discussions of the D3-brane dynamics in the throat: in particular, the Coulomb interaction between the D3-brane and anti-D3-brane can be described in this way \cite{DKM} (see also \S\ref{sec:Coulomb}).
The perturbations we will focus on in this paper are those of type (b), where the throat geometry is deformed by the coupling to the bulk compact space, or equivalently the CFT Lagrangian is perturbed by the coupling to UV physics.
The effects on the inflaton dynamics of such perturbations have
thus far been studied in only a few special cases, but we will see that they
admit a
systematic treatment.

\subsection{Leading Perturbations to the Supergravity}

We assume that the Calabi-Yau metric in (\ref{equ:warped}) can be approximated in some region by a
cone over a five-dimensional Einstein manifold $X_5$
\beq
\tilde g_{mn} \d y^m \d y^n = \d r^2 + r^2 \d s_{X_5}^2\, .
\eeq
Specifically, we have in mind the canonical example of Ref.~\cite{KS}, for which $X_5$ is the $[SU(2)\times SU(2)]/U(1)$ coset space $T^{1,1}$, and the would-be singularity at the tip of the throat, $r=0$, is smoothed by the presence of appropriate fluxes.
Roughly, this corresponds to the tip of the throat being located at a finite radial coordinate $r_{\rm IR}$, while at $r=r_{\rm UV}$ the throat is glued into an unwarped bulk geometry.
For $r_{\rm IR} \ll r < r_{\rm UV}$ the warp factor in (\ref{equ:warped}) may be written as \cite{KT}
\beq
e^{-4A(r)} = \frac{L^4}{r^4} \ln \frac{r}{r_{\rm IR}}\, , \qquad L^4 \equiv \frac{81}{8} (g_s M \alpha')^2\, ,
\eeq
where
\beq
\ln \frac{r_{\rm UV}}{r_{\rm IR}} \approx \frac{2\pi K}{3g_s M}\, .
\eeq
Here, $M$ and $K$ are integers specifying the flux background.

In this section, we discuss the leading non-normalizable perturbations to the throat that perturb the D3-brane potential and arise from coupling to the bulk Calabi-Yau moduli and to four-dimensional supergravity.  Provided that the D3-brane is well-separated from the UV region, the most important perturbations are those that diminish least rapidly toward the IR.  In the gauge theory language, these are of course the lowest-dimension operators perturbing the theory, but we will reserve the gauge theory discussion for \S3, and attempt to keep the present gravity-side discussion self-contained.\\

\noindent
{\it Harmonic analysis}
\vskip 3pt

The spectroscopy of $T^{1,1}$ was worked out in detail in \cite{Ferrara}; a nice summary is presented in Appendix A of \cite{Ofer}.  Our interest is specifically in the leading linearized perturbations of $\Phi_-$.
A systematic analysis of KK excitations around $AdS_5 \times T^{1,1}$ \cite{Gubser,Ferrara,Kim} has revealed that the modes related to $\Phi_\pm$ perturbations are
linear combinations of the conformal factor of $T^{1,1}$, $\delta g^{\, a}_a \equiv \gamma$,\footnote{Our notation can be translated to the notation of \cite{Gubser,Ferrara,Kim} by the substitutions $g \to h$, $\gamma \to \pi$.} and of the four-form perturbation with all four indices along $T^{1,1}$, $\delta C_{abcd} \equiv b$.
These modes have the harmonic expansions~\cite{Ferrara}
\beq \label{equ:gb}
\{ \gamma, b \} = \sum_{L,M} \{\gamma_{LM}, b_{LM}\} \, r^{\Delta(L)-4} \ Y_{LM}(\Psi) + c.c.\, ,
\eeq
where $\{\gamma_{LM}, b_{LM}\}$ are constants. Here $\Psi$  stands for the five angular variables, $L \equiv \{J_1,J_2, R\}$ and $M\equiv \{m_1,m_2\}$  represent the quantum numbers under the $SU(2) \times SU(2) \times U(1)_R$ global symmetry, $Y_{LM}(\Psi)$ is the corresponding angular harmonic, and
\beq \label{equ:Delta}
\Delta \equiv -2+\sqrt{6\Bigl[J_1(J_1+1)+J_2(J_2+1)-R^2/8\Bigr]+4}  \, .
\eeq
With some notational prescience, we have used the symbol $\Delta$ to represent the radial dependence; we will verify in \S3 that $\Delta$ is indeed the dimension of the corresponding operator deforming the CFT.  Here, $\Delta$ relates to the eigenvalue of the five-dimensional Laplacian on $T^{1,1}$, $\Box_5 Y_{LM} = -\Lambda Y_{LM}$, where
\beq
\Lambda \equiv 6\Bigl[J_1(J_1+1)+J_2(J_2+1)-R^2/8\Bigr]\, .
\eeq
Group-theoretic considerations give selection rules for the quantum numbers and restrict the harmonics appearing in the expansion (\ref{equ:gb}) \cite{Ferrara}.
The lowest few modes appear in Table 1 of
\cite{Ofer}; the full tower of modes is given in Table 7 of \cite{Ferrara}.\footnote{We  advise the reader that the discussion in \cite{Ofer} does not include the lowest components of current supermultiplets, which will be important for us but were not relevant for the purposes of those authors.}

Our next step is to relate the `transverse' fields $\gamma$ and $b$ appearing in (\ref{equ:gb})
 to the `longitudinal' fields $e^{4A}$ and $\alpha= C_{0123}$ that determine the D3-brane potential via $\Phi_- = e^{4A}-\alpha$.
If $ b \equiv \delta C_{abcd}\sim r^{\Delta-4}$ then $\delta F_{abcde}$ also scales as $r^{\Delta-4}$ because it contains only an additional angular derivative. Using the self-duality of the five-form field strength, and noting that the unperturbed ten-dimensional metric satisfies $\sqrt{-g}\sim r^3$, we find that $\delta F_{0123r} = \partial_r \delta C_{0123} = \partial_r \delta \alpha \sim r^{\Delta-1}$. This implies that $\delta \alpha\sim r^\Delta$.

A similar argument applies to perturbations of the warp factor.
There exists a proportionality relation between the perturbation of the conformal factor $\delta g^{\,\, \mu}_{\mu}$ and $\gamma$; for example, in the analogous case of $AdS_5\times S^5$, $\delta g^{\, \, \mu}_\mu = {16\over 15} \gamma$  \cite{Kim}.
Thus, $\delta g_{\mu\nu} \propto  \gamma \,r^2 \eta_{\mu \nu}$, so that the perturbed metric is
\begin{equation}
g_{\mu \nu}= r^2 \eta_{\mu \nu} (1+ {\rm const.} \times \gamma)= e^{2A} \eta_{\mu \nu} \ .
\end{equation}
It then follows that the perturbation to the warp factor scales as $\delta e^{4A} \sim r^4 \gamma \sim r^\Delta$.
The harmonic expansion of $\Phi_-$ may therefore be written as
\beq \label{equ:modes}
\Phi_{-}(r, \Psi) =  \sum_{L,M} \Phi_{LM}\, \Bigl(\frac{r}{r_{\rm UV}}\Bigr)^{\Delta(L)}\ Y_{LM}(\Psi) + c.c.\, ,
\eeq
where $\Phi_{LM}$ are constants.  It is straightforward to verify that $\Phi_-$ obeys the Laplace equation in the unperturbed Calabi-Yau metric of the conifold.

\vskip 8pt
\noindent
{\it Effective radial potential}
\vskip 3pt

We now digress briefly to discuss the effect of a single mode\footnote{If more than one angular mode is relevant during inflation, then the dynamics is significantly more complicated than what is described below.}
\begin{equation}\label{PhiDelta}
\Phi_-^{(\Delta)} =   \Bigl(\frac{r}{r_{\rm UV}}\Bigl)^{\Delta} \ f_L(\Psi) \, ,
\end{equation}
where
\beq
\label{equ:fL}
f_L(\Psi) \equiv \sum_M \Phi_{LM} Y_{LM}(\Psi) + c.c.
\eeq
To isolate the radial dynamics, we first minimize the potential in the angular directions. When the angular coordinates have relaxed to their minima, the potential reduces to an effective single-field potential for the radial direction $r$.
At fixed radial location, the D3-brane potential  induced by (\ref{PhiDelta}) is minimized at some angular location $\Psi_\star$ where $f_L(\Psi_\star) $ is negative.  Such an angular location always exists because any nontrivial harmonic necessarily attains both positive and negative values.\footnote{To see this, note that any non-constant harmonic is is orthogonal to the constant ($L=0$) harmonic. This implies that the integral of this harmonic has to vanish, so that the harmonic cannot be of the same sign everywhere.}
When the D3-brane sits at $\Psi_\star$, the contribution of (\ref{PhiDelta}) to the radial potential is negative, and is minimized at $r\to\infty$. This result is quite general: the potential induced by any individual perturbation of $\Phi_{-}$ produces  a radially-expulsive force.
This is fortunate, because the inflaton mass term sourced by the coupling to four-dimensional gravity provides a problematically-strong force towards the tip \cite{KKLMMT}. Inflation will be possible when these two leading forces cancel to a good approximation.

Incorporating the normalization of $\Delta V$ determined in Appendix A we may write
\begin{equation}
\label{potDelta}
\Delta V = -  c \,  a_0^4 \, T_3\,  \Bigl(\frac{\phi}{\phi_{\rm UV}} \Bigr)^{\Delta}\, ,
\end{equation}
where $\phi = \sqrt{T_3} r$ is the canonically-normalized inflaton and
$c$
is a positive constant.

To determine the leading corrections to the inflaton potential we therefore identify the nontrivial\footnote{As explained in \cite{Ofer}, the mode with $\Delta =0$ can be gauged away.} $\Phi_{-}$ modes with the smallest values of $\Delta$, {\it cf.}~equation (\ref{equ:Delta}).  Taking into account  the selection rules for $J_1$, $J_2$ and $R$, the lowest mode, $\Phi_{-}^{(3/2)}$, has $\{J_1, J_2, R\} = \{ \frac{1}{2}, \frac{1}{2}, 1\}$, yielding $\Delta=3/2$.  We also find modes $\Phi_{-}^{(2)}$ with $\{J_1, J_2, R\} = \{ 1, 0, 0\}$ and $\{0,1,0\}$, leading to $\Delta =2$.
The description of these modes in the gauge theory will be given in \S3.

There are then two generic possibilities:
\vskip 4pt
\noindent
{\bf $\mathbf{1< \Delta<2}$  (`fractional') case:}

Suppose the mode $\Phi_{-}^{(3/2)}$ is allowed to appear by all of the symmetries of the problem (see \S\ref{sec:general}).
Then the leading correction to the potential is
\begin{equation}
\label{potscale}
\Delta V = -  c \,  a_0^4 \, T_3\,  \Bigl(\frac{\phi}{\phi_{\rm UV}} \Bigr)^{3/2}\, .
\end{equation}
We have therefore reduced to the system studied in \cite{delicate}.
However, the logic here is more general; it applies even in the absence of specific embedded D7-branes in the throat geometry, and is
easily generalized to other warped throats.

\vskip 4pt
\noindent
{\bf $\mathbf{\Delta=2}$ (quadratic) case:}

As will be clear from the gauge theory analysis, it is easy to forbid the mode $\Phi_{-}^{(3/2)}$  from appearing in $\Delta V$ if
one preserves a modest discrete symmetry in the
geometry. This discrete symmetry might act in the bulk only and forbid the lowest perturbation in the throat, or it can act on the throat as well so that the throat becomes
an orbifold of the warped deformed conifold. The leading effect is then due
to $\Phi_{-}^{(2)}$
and takes the form
\begin{equation}
\label{masspert}
\Delta V = - c\,  a_0^4 T_3 \Bigl( {\phi \over \phi_{\rm UV}} \Bigr)^2\, .
\end{equation}
This mass term can be tuned against the problematic $H^2 \phi^2$ term in (\ref{basic}), and the phenomenology reduces
to that studied in \cite{KKLMMT,FT}.

\newpage
\section{Gauge/Gravity Duality for D3-brane Potentials}
\label{sec:dual}

We have seen how to compute, on the gravity side of gauge/gravity duality, the leading contributions to the inflaton potential induced by the coupling to four-dimensional supergravity and to the bulk
Calabi-Yau fields. It is very instructive to repeat this analysis on the gauge theory side.

As explained in \S2, we wish to study perturbations of the Lagrangian of the dual CFT, corresponding to {\it non-normalizable} modes in the AdS geometry.  We exclude
relevant perturbations to the four-dimensional theory, but will find that the leading corrections come from operators of the form
\begin{equation}
\label{ois}
\int d^4\theta~ M_{\rm UV}^{-\Delta} \, X^\dagger X \, {\cal O}_{\Delta}\, ,
\end{equation}
where $X$ is a moduli field that obtains a SUSY-breaking F-term vev.

Two questions immediately arise from (\ref{ois}):
\begin{enumerate}
\item Which CFT operators ${\cal O}_{\Delta}$ correspond to perturbations of $\Phi_-$ and hence induce a D3-brane potential?
\item How is the operator dimension $\Delta$ related to the radial profile of the D3-brane potential $V(r)$?
\end{enumerate}
We answer these questions in the following sections.

\subsection{Leading Perturbations to the Gauge Theory}

The operators ${\cal O}_\Delta$
that correspond to $\Phi_-$ perturbations
are those that produce a potential on the Coulomb branch of the CFT; hence, they should be made out of the scalar fields that parametrize this Coulomb branch.
To enumerate the lowest-dimension such operators one can begin by listing chiral and otherwise protected operators in the gauge theory, with the understanding that at strong 't Hooft coupling a gravity-side analysis is necessary to determine the dimensions of more general operators.

The gauge theory dual to the warped deformed conifold geometry is an $SU(N+M) \times SU(N)$
gauge theory with bi-fundamental fields $A_i, B_j$ ($i,j = 1,2$).
The single-trace operators built out of these scalar fields and their complex conjugates are labeled by their $SU(2)_A\times SU(2)_B\times U(1)_R$ quantum numbers $(J_1, J_2, R)$.
 Using the AdS/CFT correspondence, the dimensions of these operators are given by (\ref{equ:Delta}), though in the highly protected cases of interest to us, the dimensions can be determined directly in the gauge theory as well.

\vskip 6pt
\newpage
\noindent
{\it Chiral operators}
\vskip 3pt

For $J_1= J_2= R/2$, these operators are chiral and have the simple form
\be\label{chiral}
{\cal O}_{\Delta} = {\rm Tr} \Bigl( A^{(i_1} B_{(j_1} A^{i_2} B_{j_2} \ldots A^{i_R)} B_{j_R)} \Bigr) + c.c.
\ee
The dimensions of these chiral operators, $\Delta= 3R/2$, are fixed by the ${\cal N}=1$ superconformal invariance.
The lowest-dimension such operators are
\be
\label{equ:chiral}
{\cal O}_{3/2} = {\rm Tr} \left( A_i B_j\right) + c.c.\, ,
 \ee
 which have $\{J_1, J_2, R\} = \{\frac{1}{2}, \frac{1}{2}, 1\}$. These chiral operators have $\Delta=3/2$ and determine the leading term in the inflaton potential via (\ref{ois}), unless they are forbidden to appear by symmetries that are preserved by the full string compactification; see \S4.

\vskip 6pt
\noindent
{\it Non-chiral operators}
\vskip 3pt

 There are a number of operators which have the next lowest dimension, $\Delta =2$. For example, there are operators with $\{J_1, J_2, R\} = \{1, 0, 0\}$:\footnote{An additional operator with $\Delta = 2$ is the
$SU(2) \times SU(2) \times U(1)_R$ singlet operator $U$ that belongs to the baryon number current multiplet. This operator is responsible for resolution of the conifold \cite{KW2}, and it sources a D3-brane potential in the throat at the non-linear level \cite{DKS}.}

 \be \label{dimtwo}
{\cal O}_2  = \quad {\rm Tr}\left( A_1 \bar A_2\right)\ , \quad {\rm Tr} \left(A_2 \bar A_1\right)\ , \quad {1\over \sqrt 2} {\rm Tr} \left(A_1\bar A_1 - A_2 \bar A_2\right)
 \ ,
 \ee
 and the corresponding $\{J_1, J_2, R\} = \{0, 1, 0\}$ operators made out of the fields $B_j$. While non-chiral, these operators are protected because
 they are related by supersymmetry to $SU(2)\times SU(2)$ currents; therefore, their dimension, 2, is exact in the gauge theory.\footnote{See footnote \ref{foot}.}
 Using gauge/gravity analysis we can see
 that the above operators source an inflaton potential at the linearized level; therefore, they will play an important role in our considerations.  In particular, if the chiral operators of dimension $\Delta=3/2$ are forbidden by discrete symmetries, the non-chiral
 operators (\ref{dimtwo}) will determine the leading corrections due to bulk effects, and will allow a direct tuning of the
 inflaton mass.

\subsection{D3-brane Potential via AdS/CFT}

We now comment on the correspondence between operator dimensions in the CFT and the radial profiles of bulk contributions to the D3-brane potential.
Let us consider a conformal gauge theory on a stack of D3-branes perturbed by an operator ${\cal O}_{\Delta}$ of dimension $\Delta$. If the gauge theory has a Coulomb branch corresponding to separating a D3-brane from the stack, then we expect such an operator to produce a potential scaling as $\phi^\Delta$ where $\phi$ is the scalar field corresponding to the D3-brane position.
In gauge theories that have near-AdS gravity duals we can demonstrate this result, for a class of operators, by
 studying small perturbations to the background that are dual to the operators ${\cal O}_\Delta$.

\vskip 6pt
\noindent
{\it Operator dimensions and radial scaling of $\Delta V$}
\vskip 3pt

Let us now add a new entry to the AdS/CFT dictionary by considering the behavior of the D3-brane potential on the Coulomb branch.
Our study will be rather general, although we will mainly use  $AdS_5\times T^{1,1}$ as an example.
 We will utilize the well-known correspondence \cite{ADSCFT} between the dimensions of gauge-invariant operators and the radial scaling of bulk AdS fields: in terms of the $AdS_5$ metric $\dd s^2_{AdS} = r^2 \dd x^\mu \dd x_\mu + \dd r^2/r^2$, the solutions for a field $\chi$ associated with an operator of dimension $\Delta$ are
\beq
\label{equ:radial}
\chi(r) = \chi_1 r^{-\Delta} + \chi_0 r^{\Delta -4}\, .
\eeq
Here $\chi_0$ is dual to the source for the operator ${\cal O}_\Delta$, $\chi_1$ is dual to its expectation value \cite{KW2}, and $r$ is dual to the energy scale in the gauge theory. If we introduce a UV cutoff  $r_{\rm UV}$, then $\chi_0$ scales as $r_{\rm UV}^{4-\Delta}$, while $\chi_1$ scales as $r_{\rm UV}^\Delta$.

Comparing (\ref{equ:radial}) to (\ref{equ:gb}) and (\ref{equ:modes}) we see that $\Delta$ as given in (\ref{equ:Delta}) is in fact the dimension of the operator dual to the associated perturbation of $\Phi_-$, justifying our choice of notation in \S2.
Importantly, we have established the following correspondence between the operator dimension and the radial profile of the D3-brane potential,
\be
\Delta V \propto r^\Delta\, ,
\ee
in agreement with the field theory expectations for the scaling of the potential on the Coulomb branch.
As is obvious from our derivation, this result is general to the AdS/CFT correspondence and applies to any $AdS_5\times X_5$ background of type IIB string theory.

\vskip 6pt
\noindent
{\it Chiral operators}
\vskip 3pt

If we perturb the theory by a chiral operator (\ref{chiral}), then the potential is found to behave in accordance with classical expectations. Namely, if we substitute for $A_i,B_j$ the classical formulae \cite{IgorRemarks}
\bea \label{classform}
A_1 &\sim & r^{3/4} \sin  (\theta_1/2) e^{i (\psi - \phi_1)/2} \ , \qquad
 A_2 \sim r^{3/4} \cos (\theta_1/2) e^{i (\psi + \phi_1)/2}\ , \nonumber \\
 B_1 &\sim & r^{3/4} \sin (\theta_2/2) e^{i (\psi - \phi_2)/2} \ , \qquad
 B_2 \sim r^{3/4} \cos (\theta_2/2) e^{i (\psi + \phi_2)/2}
\ ,\eea
we obtain the correct dependence of the potential on $r$ and on the angular directions of $T^{1,1}$ even at strong coupling, as confirmed by our dual gravity calculation:
\be
\label{equ:VV}
\Delta V\sim - c\, a_0^4 \, T_3\, \Bigl(\frac{r}{r_{\rm UV}}  \Bigr)^{3R/2} Y_{LM}(\Psi) + c.c.
\ ,
\ee
where $L = \{\frac{R}{2}, \frac{R}{2}, R\}$.
 We explain the overall normalization of (\ref{equ:VV}) in Appendix A.

\vskip 6pt
\noindent
{\it Non-chiral operators}
\vskip 3pt

Not surprisingly, this simple classical argument  is not applicable for generic non-chiral operators that contain $\bar A$ and/or $\bar B$; this is in part because such operators obtain generally irrational anomalous dimensions at strong 't Hooft coupling, where the gravity description is valid.
An important special case is provided by the lowest components of supermultiplets of currents, which are protected.
We will see that in some circumstances these non-chiral operators in fact give the leading contribution to the inflaton potential; see \S4.

\subsection{Remarks on the Coulomb Potential}
\label{sec:Coulomb}

For completeness, we now describe the Coulomb interaction between a D3-brane in the throat and an anti-D3-brane at the tip of this throat, on both the gravity side and the gauge theory side.

The Coulomb potential can be calculated by viewing the mobile D3-brane as a perturbation to $\Phi_+ = 2 e^{4A}$. The effect of a D3-brane on the warp factor $e^{-4A}$ was calculated in \cite{Klebanov:2007us}; inverting this result we find
\begin{equation}
\label{phipluspert}
V_{D3/\overline{D3} }(r) = T_3\, \Phi_+(r; r_0) = 2 a_0^4 T_3 + a_0^4 T_3 \left[\sum_{L,M} \Phi_{LM} \left(\frac{r_0}{r}\right)^{4+ \Delta(L)} Y_{LM}(\Psi) + c.c.\right]\, ,
\end{equation}
where $a_0 \equiv e^{A(r_0)}\sim r_0/r_{\rm UV}$, while $r_0$ represents the radial coordinate in the throat where the anti-D3-brane is located.
This perturbation introduces a correction to the anti-D3-brane energy whose magnitude depends on the D3-brane position $r$, and therefore induces a potential for the D3-brane \cite{KKLMMT}.
The leading term, corresponding to $\Delta=0$, is the usual radial Coulomb potential. As noted in \cite{Kachru:2007xp}, this term corresponds to adding the dimension-8 operator ${\rm{Tr}} (F^4)$ in the dual gauge theory.
The remaining terms give subleading, angular-dependent corrections corresponding to operators of yet higher dimension. 

The formula (\ref{phipluspert}) for perturbations of $\Phi_+$ is valid for any $r_0 < r$. After introducing a factor $r_0^{-4}$ to transform to perturbations of the fields on $T^{1,1}$ (see
  the discussion leading up to (\ref{equ:modes})), we observe that these perturbations grow with $r_0$ as $r_0^{4+ \Delta (L)}$. This means that in the infrared gauge theory these perturbations correspond to adding sources for operators of dimension $8+\Delta (L)$. The corresponding operators have the schematic form
${\rm{Tr}} (F^4 A_i B_j \ldots)$.

Alternatively, one can arrive at the same result by solving (\ref{equ:PhiEoM}) for the normalizable $\Phi_-$ profile created by
an anti-D3-brane; such a perturbation directly introduces a potential for D3-brane motion, $V_{D3/\overline{D3}}=T_3 \, \Phi_-$.
Reading off $\Phi_-(r)$ from (\ref{phipluspert}) and multiplying by $r^{-4}$ to transform to the ``transverse'' field components, we find
that their perturbations fall off as $r^{-8-\Delta(L)}$. It follows that these perturbations correspond to expectation values for operators of dimension $8+\Delta(L)$ like ${\rm{Tr}} (F^4 A_i B_j \ldots)$.
Thus, either way of interpreting the Coulomb potential leads to the conclusion that it is mediated by the above gauge theory operators.  However, note that the D3-brane introduces a non-normalizable $\Phi_+$ perturbation, or equivalently introduces a {\it{source}} for these operators, whereas the anti-D3-brane induces a normalizable $\Phi_-$ perturbation, or equivalently leads to an {\it{expectation value}} for these operators.

The normalizable perturbations $\delta \Phi_+$ correspond to different operators.
Perturbations of the warp factor satisfy
$\tilde \nabla^2 (\delta e^{-4A}) =0$.
The normalizable fluctuations \cite{Klebanov:2007us}
\begin{equation}
\delta e^{-4A} \sim r^{-4} \left(\frac{r_0}{r} \right)^\Delta Y_{LM}(\Psi)
\end{equation}
when plugged into the angular metric $e^{-2A} r^2 \d s^2_{T^{1,1}}$ then give metric perturbations $\sim r^{-\Delta} Y_{LM}$.
These correspond to vevs of operators of dimension $\Delta$
like ${\rm{Tr}}(A_i B_j)$. In \S2.2 we also concluded that sources for these operators correspond to non-normalizable fluctuations of $\Phi_-$.

Thus, by studying both normalizable and non-normalizable perturbations $\delta \Phi_+$ and $\delta \Phi_-$ we are led to the following conclusions:
\begin{enumerate}
\item[i)] For perturbations $\delta \Phi_-$ the non-normalizable modes correspond to sources of 
%scalar 
operators of dimension $\Delta$ like ${\rm{Tr}}(A_i B_j)$,
but normalizable perturbations correspond to vevs of operators of dimension $8+\Delta$ such as ${\rm{Tr}} (F^4 A_i B_j)$.
\item[ii)] For perturbations $\delta \Phi_+$
the non-normalizable modes correspond to sources of operators of dimension
$8+\Delta$ like ${\rm{Tr}} (F^4 A_i B_j)$, but normalizable perturbations correspond to vevs of operators of dimension $\Delta$ such as
${\rm{Tr}}(A_i B_j)$.
\end{enumerate}
While this was implicit in earlier literature, we believe that this curious asymmetry has not been emphasized.

\subsection{Revisiting the Eta Problem}

Having given a classification of the leading corrections to the inflaton potential from bulk effects, we now revisit the origin of the $H^2 \phi^2$ term in (\ref{basic}).  In the presence of
SUSY-breaking by a moduli F-term $F_{X}$, one generically expects contributions to the inflaton potential of the form
\begin{equation}
\label{etaprob}
\int  d^4 \theta ~M_{\rm UV}^{-2} \, X^\dagger X\,  {\cal K}_{\rm CFT}
\end{equation}
where ${\cal K}_{\rm CFT}$ is the K\"ahler potential of the CFT.
This is the contribution responsible for the famous supergravity eta problem \cite{copeland}, and was found directly in
\cite{KKLMMT} for the warped D-brane model of interest to us.

How do we understand the appearance of such a term, in the context of the approach presented in this paper?  We note the potentially confusing fact
that in the classification of \cite{Ferrara}, there are no operators ${\cal O}_{\rm CFT}$ which have $\Delta =2$, are singlets
under the global symmetries of the CFT, and correspond to $\Phi_-$
perturbations.  Hence, it may naively seem difficult to generate the appropriate $\phi^2$ terms in the potential, by
UV perturbations of the form
(\ref{etaprob}).

However, there is no problem in identifying such a perturbation on the gravity side.
In fact, this is precisely the term that was computed in \cite{KKLMMT} by studying a probe D3-brane in a warped flux background with nonperturbative
moduli stabilization.  %But
So why then do we find no corresponding $\Delta =2$ operator in the CFT?

The general philosophy of the renormalization group (RG), as applied to the conifold CFT, is as follows.   The tree-level K\"ahler potential
that gives rise to the correct Coulomb branch metric (as seen by probe D3-branes) is
\begin{equation}
\label{classical}
{\cal K}_{\rm classical} \sim \left({\rm Tr}\sum_{i,j=1}^2 A_i^\dagger A_i B_j B_j^\dagger  \right)^{2/3} .
\end{equation}
Quantum mechanically in the large 't Hooft coupling limit, this operator acquires a very large dimension $\Delta \sim (gN)^{1/4}$; it is
dual to a string state, not a supergravity mode.  However, we know that the RG flow of the theory with classical K\"ahler potential
(\ref{classical}) must be such that at the IR critical point, there exists some ${\cal K}_{\rm CFT}$ that preserves scale invariance.
The absence of a marginal perturbation to this ${\cal K}_{\rm CFT}$ (as indicated by the spectroscopy of \cite{Ferrara}, or by direct counting of moduli
in the closed string dual) is not in conflict with the existence of an admittedly uncomputable ${\cal K}_{\rm CFT}$ at the fixed point.  We note that this is not a peculiar feature of the conifold CFT; it is a common phenomenon, exhibited by
{\it e.g.}~supersymmetric QCD theories in the conformal window, supersymmetric minimal models, and in fact most superconformal field theories known to us.
In sum, there is no inconsistency between a) the presence of an inflaton mass generated by coupling to
supergravity, and b) the absence in the CFT of a $\Delta=2$ operator that is dual to a $\Phi_-$ perturbation and is invariant under all global
symmetries.

\section{The Inflaton Mass and Discrete Symmetries}
\label{sec:general}

We have described a general framework for relating the leading terms in the potential for a D3-brane in a warped throat to perturbations of the dual gauge theory Lagrangian by low-dimension operators.  Clearly, a central question in this approach is the spectrum of operator dimensions in the CFT.  In \S\ref{sec:unbroken}  we explain the role of discrete symmetries preserved by the UV physics in determining the lowest-dimension operator perturbations induced by the compactification.  In \S\ref{sec:orbifold} we remark that certain classes of CFTs manage to have no chiral operators with $\Delta<2$, even before the coupling to UV physics.

\subsection{Unbroken Bulk Symmetries}
\label{sec:unbroken}

In {\it{known}} examples of CFTs with gravity duals, continuous global symmetries of the CFT, corresponding to isometries of the throat geometry, are always present.  Compact Calabi-Yau spaces admit no continuous isometries, however, so the coupling of the throat to the compact bulk necessarily breaks the throat isometries down to a (possibly trivial) discrete subgroup.  We refer to these residual symmetries as unbroken discrete symmetries.

Our general strategy has been to perturb the CFT Lagrangian by the most general operators consistent with all symmetries of the problem; in particular, no perturbation that is forbidden by unbroken discrete symmetries will be turned on.  Chiral operators are readily forbidden in this way. For example,
 ${\rm Tr}(A_i B_j)$ is forbidden from appearing by {\it e.g.}~a ${\mathbb{Z}}_2$ symmetry acting as $A_i \to -A_i$, and it is straightforward to arrange for the full compactification to preserve such a discrete subgroup of the global symmetry of the CFT.  Perturbations by non-chiral operators, however, are much more difficult to forbid; and as we have seen, $\Delta=2$ non-chiral perturbations in the supermultiplets of conserved currents are present in any CFT with a continuous global symmetry.

 Thus, the chiral ${\cal O}_{3/2}$ perturbations lead to the dominant contribution to the inflaton potential unless they are forbidden by unbroken discrete symmetries.
 In contrast, the non-chiral ${\cal{O}}_2$ perturbations are generically present.

\subsection{Chiral CFTs}
\label{sec:orbifold}

The above discussion applies to the conifold CFT, and more generally to any CFT possessing a chiral operator with $1< \Delta<2$.  However, the presence of such operators
is certainly not a universal feature of the large $N$ CFTs that have
known gravity duals.  For instance, in ${\cal N}=4$ supersymmetric
Yang-Mills theory, the lowest-dimension operators (chiral and
non-chiral) have $\Delta = 2$.  The same applies to many orbifolds of this
theory by subgroups of its $R$-symmetry group \cite{ShamitEva,LNV}; for instance,
the ${\cal N}=2$ supersymmetric theory arising from $N$ D3-branes on
$\mathbb{C}^2/\mathbb{Z}_2$ again has (chiral and non-chiral) operators with $\Delta = 2$
as its lowest-dimension perturbations.  Below, we describe some CFTs in
which the lowest-dimension operators are non-chiral operators in the
supermultiplets of currents and have $\Delta = 2$; we focus on these
examples because cascading geometries based on small perturbations of
these theories have already been studied in the literature.

%\vskip 25pt
\newpage
\noindent
{\it $\mathbb{Z}_2$ orbifold of the conifold}
\vskip 3pt

One well-known example of a theory where $\Delta < 2$ operators do
not appear is the ${\mathbb Z}_2$ orbifold of the warped deformed conifold by $z_a\rightarrow
- z_a$, $a=1,\ldots 4$. This warped throat metric is the same as in
\cite{KS}, but the range of the $\psi$ coordinate is reduced from
$[0,4\pi)$ to $[0,2\pi)$. From the point of view of the cascading
gauge theory, the ${\mathbb Z}_2$ action is best thought of as \be
\label{zetwo} A_i\rightarrow - A_i\ , \quad i=1,2 \, .\ee This
produces a chiral $[SU(N+M)\times SU(N)]^2$ gauge theory whose
duality cascades were studied in {\it e.g.}~\cite{Franco}.

Up to logarithmic corrections, the asymptotic geometry is that of
$AdS_5\times T^{1,1}/{\mathbb Z}_2$, where the ${\mathbb Z}_2$ acts freely. This configuration is
created by D3-branes at the tip of a complex cone over $\mathbb{F}_0= \mathbb{P}_1 \times \mathbb{P}_1$ \cite{Morrison}. In the dual chiral CFT, $\Delta=3/2$ chiral operators (\ref{equ:chiral}) are removed
by the ${\mathbb Z}_2$ projection (\ref{zetwo}). The lowest-dimension
operators in this theory are $\Delta=2$ operators of the form
(\ref{dimtwo}), and the corresponding operators made out of the fields $B_j$, that belong to the
$SU(2)\times SU(2)$ current multiplets.\footnote{We remark that, in the $AdS_5\times
T^{1,1}/{\mathbb Z}_2$ approximation, the associated contribution to the potential actually scales as $r^2 \ln
r$ (corresponding to a source for an operator with $\Delta=2$
\cite{KW2}), so at very large $r$ the repulsion will dominate over the attractive force of \cite{KKLMMT}. }

\vskip 6pt
\noindent
{\it $Y^{p,q}$ throats}
\vskip 3pt

The identification of the lowest-dimension operators with the leading corrections to the inflaton potential
has broad implications for models based on arbitrary
warped throats, not just the warped deformed conifold.

As we have seen above, the fact that the ${\mathbb Z}_2$-orbifolded conifold
gauge theory is chiral implies that the chiral operators of $\Delta < 2$ are absent
from the spectrum. This feature of chiral gauge theories seems to
be quite generic. For instance, it appears to extend to the entire class of
theories dual to $AdS_5\times Y^{p,q}$, where $ p>q$. The chiral
ring in these theories was studied in \cite{Berenstein,Kihara} with
the conclusion that the lowest dimension chiral operator has
either dimension $3$, or dimension \be \frac{3}{2} \left(p-q+ s \right) \, ,
\qquad {\rm or}\qquad \frac{3}{2} \left (p+q- s \right) \, ,
\ee where $s \equiv (3q^2- 2 p^2 +
p\sqrt{4p^2-3q^2})/(3q)$. It is not hard to check that these dimensions
exceed $2$ for all $p>q$. As a result, the lowest-dimension scalar
operators that are ${\it known}$ have dimension 2 and belong to current multiplets; such current multiplets are always present, as all $Y^{p,q}$ spaces have
$SU(2)\times U(1)\times U(1)_R$ symmetry.
The full KK spectrum of these models is not known, and so it is possible that there
exist non-chiral operators with $ \Delta < 2$, but this seems unlikely based on
experience with simpler models.
Thus, the phenomenology of
inflation models in throats based on $Y^{p,q}$ models \cite{HEK}
is likely quite similar to that in the ${\mathbb Z}_2$ orbifold of the warped deformed conifold
discussed above.

\section{Phenomenology and a Classification of Models}

In this paper we have given a systematic classification of %all
%DB
the leading
 contributions to the D3-brane potential in warped throat geometries and the associated contributions to the slow-roll eta parameter
\begin{eqnarray}
\label{equ:VV}
V(\phi) &=& V_0(\phi) \ \ + \ \ H_0^2 \phi^2 \ \ +  \ \ \Delta V(\phi) \\
\eta(\phi) &=& \ \ \, \eta_0 \ \ \  \,\, +  \  \ \ \ \frac{2}{3} \ \ \ \ \, + \ \ \, \Delta \eta(\phi)  \ \ =  \ \ ?
\end{eqnarray}
Here, $V_0(\phi)$ is {\it defined} to be all terms in $V(\phi)$ with negligible contributions to the eta parameter, {\it i.e.}~$\eta_0 \ll 1$.
This includes the brane-antibrane Coulomb interaction
\beq
V_{D3/\overline{D3}}(\phi) = D \left( 1- \frac{3 D}{16 \pi^2} \frac{1}{\phi^4}\right)\, , \qquad D \equiv 2 a_0^4 T_3\, ,
\eeq
as well as any other sources of nearly-constant energy, {\it e.g.}~bulk contributions to the cosmological constant.
The mass term in (\ref{equ:VV}), with
\beq
3 M_{\rm pl}^2 H^2_0 \equiv V_0(\phi_{\rm UV}) \, ,
\eeq
generically makes the potential too steep for prolonged inflation to occur.
The eta parameter can only be small if the additional contributions to the potential $\Delta V$
can cancel this mass term.

In this paper we have given a framework in which the contributions to $\Delta V$ can be classified and their effects on the inflationary dynamics systematically analyzed.
In particular, we have found the following spectrum of correction terms to the inflaton potential
\beq
\label{equ:potentialAngles}
\Delta V (\phi, \Psi) =  \sum_{L,M} c_\Delta  \left(\frac{\phi}{\phi_{\rm UV}}\right)^{\Delta(L)} \alpha_{LM} Y_{LM}(\Psi) +c.c.\, ,
\eeq
where
the constants $c_\Delta$ are strongly model-dependent (see Appendix A).
In general, the potential (\ref{equ:potentialAngles}) can have a complicated angular dependence.
However, if only one angular mode dominates, the total potential (\ref{equ:VV}) reduces to
\beq \label{equ:potential}
V (\phi)~=~V_{0}(\phi)  ~+~ M_{\rm pl}^2 H^2_0 \left[ \left(\frac{\phi}{M_{\rm pl}} \right)^2 - c_\Delta \, \left(\frac{\phi}{\phi_{\rm UV}}\right)^\Delta \right] \, .
\eeq
In the following we restrict ourselves to a discussion of (\ref{equ:potential}) and leave a more detailed analysis of (\ref{equ:potentialAngles}) for a future study.
The lowest-dimension operator allowed by the symmetries of the compactified throat geometry then determines the inflationary phenomenology associated with the potential (\ref{equ:potential}).

\subsection{Summary: A Classification of Models}

We may separate possible D-brane inflation models based
on gauge/string duality into three classes:

\begin{enumerate}
\item[i)] Fractional Case: $1< \Delta < 2$

In the first class, the CFT coupled to the Calabi-Yau sector contains operators of dimension $1< \Delta<2$ that produce a potential on the Coulomb branch.  Clearly, this is possible only if the CFT itself contains such operators, and perturbations by these operators are allowed by any (discrete) global symmetries preserved by the bulk.  The generic form of the inflaton potential is then (\ref{equ:potential}) with $1<\Delta< 2$.
In this case, we expect that fine-tuned inflation near an approximate inflection point is typically possible. We describe the phenomenology of this case in \S\ref{sec:fractional}.

\item[ii)] Quadratic Case: $\Delta = 2$

In the second class, the lowest-dimension operators allowed by the global symmetries of the bulk have
$\Delta=2$.
This is the generic situation in warped throats dual
to chiral gauge theories.
In \S\ref{sec:general} we explained how even in non-chiral gauge theories, simple discrete symmetries can forbid any $\Delta<2$
chiral operators while allowing operators with $\Delta=2$.
 In this case the potential takes the form
\beq \label{equ: secondcase}
V (\phi)~=~V_{0}(\phi) ~+~  \beta H^2_0  \phi^2 \, ,
\eeq
where the parameter $\beta$ can be fine-tuned by varying the strength of the UV perturbation.
One can therefore obtain models of small-field inflation similar to those considered in \cite{KKLMMT,FT}.
We comment on the phenomenology of this case in \S\ref{sec:quadratic}.

\item[iii)] Unworkable Case: $\Delta > 2$

Finally, we could imagine warped throats dual to gauge theories
where all contributing operators have dimensions $ \Delta >2$. If such throats are
supersymmetric, they cannot have continuous isometries,
since the latter would imply the presence of $ \Delta =2$ scalar
superpartners of conserved currents.  Constructing such throats is technically challenging and at present no examples exist. In any event, such throats
are not desirable for brane inflation, because the leading correction to the inflaton potential is generically too small to solve the eta
problem.
\end{enumerate}

\subsection{Comments on Phenomenology}

\addtocontents{toc}{\SkipTocEntry}
\subsubsection{`Fractional' Case: $\mathbf{1<\Delta < 2}$}
\label{sec:fractional}

If the symmetries of a compactification of the warped deformed conifold allow a perturbation by the chiral operator $ {\rm Tr}(A_i B_j)$ with $\Delta = 3/2$,
we obtain the following phenomenological potential for the inflaton\footnote{As explained in \cite{delicate}, it is crucial to include the Coulomb interaction between the brane and the antibrane in order to get a monotonic potential.}
\beq \label{equ:pot}
V (\phi)~=~V_{0}(\phi)  ~+~ M_{\rm pl}^2 H^2_0 \left[ \left(\frac{\phi}{M_{\rm pl}} \right)^2 - a_{3/2} \, \left(\frac{\phi}{M_{\rm pl}}\right)^{3/2} \right] \, ,
\eeq
where $a_{3/2} \equiv c_{3/2} \left(\frac{M_{\rm pl}}{M_{\rm UV}}\right)^{3/2} \sim {\cal O}(1)$ (see Appendix A).
This is %identical
%DB
similar to the functional form discussed in \S4 of \cite{delicate}, but the microscopic interpretation of the $\phi^{3/2}$ term is now different.  This implies that the microscopic constraints on the coefficient $a_{3/2}$ that were important in the context of \cite{delicate} will not apply, and the corresponding difficulties in fine-tuning the inflationary potential are significantly reduced.

The slow-roll parameter $\eta$ corresponding to (\ref{equ:pot}) is
\beq
\eta(\phi) = \frac{2}{3} - \frac{3 a_{3/2}}{4}  \left(\frac{M_{\rm pl}}{\phi}\right)^{1/2} ~+~ \eta_{0}(\phi)\, .
\eeq
As in \cite{delicate} we notice that $\eta$ is large and negative for small $\phi$, {\it i.e.}~$\lim_{\phi \to 0} \eta= - \infty$.
To have inflation near some location $\phi_\star$ inside the throat
($\phi_\star < \phi_{\rm UV} \sim M_{\rm UV}$) we require
$\eta(\phi_\star) \approx 0$, and hence, using
$\eta_{0}(\phi_{\rm UV}) \ll 1$,
\beq
\eta (\phi_{\rm UV}) \approx \frac{2}{3}  -\frac{3 c_{3/2}}{4} \left( \frac{M_{\rm pl}}{\phi_{\rm UV}} \right)^2 > 0
\eeq
or
\beq
\label{equ:finetuning}
c_{3/2} < \frac{8}{9} \left( \frac{\phi_{\rm UV}}{M_{\rm pl}} \right)^2 \lesssim \frac{4}{ N}\, .
\eeq
In the second inequality of (\ref{equ:finetuning}) we have applied the field range bound of \cite{BM}.
Since naturally $c_{3/2} \lesssim {\cal O}(1)$, the fine-tuning required to get `inflection point inflation' inside the throat, $c_{3/2} < 4/N$, seems moderate.\footnote{In particular, it seems to alleviate the delicate fine-tuning found in \cite{delicate} and further explored in \cite{Panda}.  In fact, even in the setting of \cite{delicate}, there are further effects that appear to make the fine-tuning less delicate
\cite{Underwood:2008dh}, \cite{Hoi:2008gc}.}
Moreover, as explained in Appendix A, there is a straightforward physical interpretation of suppressing the value of $c_{3/2}$: one reduces the F-term potential of a D3-brane in the UV region by arranging that all moduli-stabilizing divisors are far from the inflationary throat.

\addtocontents{toc}{\SkipTocEntry}
\subsubsection{Quadratic Case: $\mathbf{\Delta = 2}$}
\label{sec:quadratic}

If the leading operator has dimension $\Delta =2$, the phenomenology is quite different.
The potential (\ref{equ:potential}) takes the form
\beq
V(\phi) = V_{0}(\phi) + H^2_0 \underbrace{\left[1- c_2 \left( \frac{M_{\rm pl}}{M_{\rm UV}}\right)^2 \right]}_{\equiv \ \beta} \ \phi^2\, ,
\eeq
with a tunable $\Delta V \sim \beta H^2_0  \phi^2$ term.  The phenomenology of warped brane inflation with a mass that is fine-tuned to solve the eta problem was
discussed for $\beta \ll 1$ in \cite{KKLMMT} and parametrized in detail for $\beta \lesssim 1$ in \cite{FT, Bean:2007hc}.

It is worth noting that in the quadratic case
the potential can be
flattened in a much larger field range than in the inflection point models.
Nevertheless, one cannot obtain monomial large-field inflation, even if this were consistent with the form of the potential, because of the field-range bound of \cite{BM}.

Finally, let us remark that an expulsive potential is required for a DBI inflation scenario \cite{DBI}
in which the brane moves out of the throat \cite{IR, BeanIR}.
It would be interesting to use our results to parametrize possible models in this class.

\section{Conclusions}

We have presented a systematic approach to enumerating possible
contributions to the potential of a D3-brane in a warped throat region
of a compact Calabi-Yau space.  The effects of
supersymmetry-breaking deformations in the IR region of the
throat, corresponding to perturbations of the state of the dual
(approximate) CFT, can be computed directly in any well-specified
example -- {\it e.g.}, one can easily compute the Coulomb potential produced
by an anti-D3-brane -- and we did not advance this subject in the
present work.  On the other hand, supersymmetry-breaking deformations in
the UV, corresponding to deformations of the Lagrangian of the dual CFT,
will in general make important contributions to the D3-brane potential,
in a manner that appears {\it a priori} to be unconstrained and to depend
sensitively on the details of moduli stabilization and of the gluing of
the throat into the compact space.

Our strategy was to recognize that the leading contributions to the
potential  that arise from UV deformations correspond to perturbations by the
lowest-dimension operators in the dual CFT that generate a potential on
the Coulomb branch.  Equivalently, these are the perturbations of the
supergravity mode $\Phi_{-}$ (\ref{equ:phim}) with the smallest possible eigenvalues of
the angular Laplacian.  Although the {\it  coefficients}  of these
perturbations do depend sensitively on the properties of the compact
space, the {\it form} of the resulting D3-brane potential is dictated by
the structure of the CFT.

Using existing results \cite{Ferrara} on the Kaluza-Klein spectrum of
$T^{1,1}$, we gave an example of our approach, exhibiting the leading
corrections to the inflaton potential for the case of the warped
deformed conifold \cite{KS}.  The lowest-dimension contributing operator
is chiral, with $\Delta = 3/2$; when not forbidden by discrete symmetries, it generates a
$\phi^{3/2}$ term in the inflaton potential.  The phenomenology is then
%identical
similar
%DB
to that which arises when moduli-stabilizing D7-branes enter
the throat region \cite{delicate}, but occurs far more generally.  The
next contribution, which is the dominant effect in the presence of
discrete symmetries that forbid the leading chiral term, comes from a
non-chiral operator of dimension $\Delta = 2$, related by supersymmetry to
a global symmetry current.  This perturbation introduces a negative
contribution to the inflaton mass-squared: it creates an expulsive force
on the D3-brane.  This can be balanced against the problematically-large
and positive inflaton mass described in \cite{KKLMMT}, and by
fine-tuning the coefficient of the perturbation, it seems that the
phenomenology conjectured in \cite{KKLMMT} and studied in \cite{FT} can
be obtained in a consistent construction.

Our methods extend immediately to more general throat geometries, given
sufficient knowledge of the Kaluza-Klein spectrum, or equivalently of
the
spectrum of lowest-dimension operators.  In
theories with global symmetries, it appears that fine-tuning the
inflaton mass to obtain inflation is straightforward.  We
remark that in theories without global symmetries and in which the
lowest-dimension chiral operators have $\Delta>2$, fine-tuned warped
brane inflation is apparently impossible.  At present, we do not have
examples of such throats.

\newpage
\section*{Acknowledgements}

We are grateful to O.~DeWolfe, L.~Kofman, J.~Maldacena, and M.~Mulligan for useful discussions.
The research of D.B. is supported in part by the David and Lucile Packard Foundation and the Alfred P.~Sloan Foundation and by Fellowships of the Center for the Fundamental Laws of Nature and the Center for Astrophysics at Harvard.  A.D. and S.K. are supported by the Stanford Institute for Theoretical Physics, the NSF under grant PHY-0756174, and the DOE under contract DE-AC03-76SF00515. The research of A.D. is also supported in part by grant RFBR 07-02-00878, and Grant for Support of Scientific Schools NSh-3035.2008.2.  A.D. would like to thank the Galileo Galilei Institute for Theoretical Physics, where part of this work was done, for hospitality.  S.K. is grateful to the
Kavli Institute for Theoretical Physics, the
Aspen Center for Physics, and the Institute for Advanced Study for hospitality while some of these ideas were being finalized.
The research of I.R.K. was supported in part by the NSF under grant PHY-0756966.
I.R.K. thanks
the IHES for hospitality during the final stages of this work.    The research of L.M. is supported by NSF grant PHY-0355005.
L.M. thanks the Stanford
Institute for Theoretical Physics for hospitality while some of this work was performed, and the high energy theory group at Harvard for hospitality while it was finalized.

\newpage
\appendix

\section{The Scale of Moduli F-terms}
\label{sec:Fterm}

In this appendix we explain the origin, and estimate the scale, of the F-term potential experienced by D3-branes in a nonperturbatively-stabilized conformally-Calabi-Yau flux compactification.

As we have reviewed, a D3-brane in a compactification with imaginary self-dual fluxes \cite{Giddings} experiences no classical force.  However, incorporating nonperturbative effects that stabilize the K\"ahler moduli will also generically lift the D3-brane moduli space \cite{KKLMMT,DeWolfe}. To see this concretely, consider the moduli stabilization scenario of \cite{KKLT} in the case of a single K\"ahler modulus, $\rho$, governed by the superpotential \cite{GVW} and K\"ahler potential \cite{deWG}\footnote{For recent studies of the K\"ahler potential in warped compactifications, see \cite{Douglas}.}
\beq \label{equ:wis}
  W=\int G_3\wedge\Omega + A(y)e^{-a\rho}  \, , \qquad
       {\cal  K}=-3~{\rm log}\Bigl(\rho+\bar\rho-k(y,\bar{y})\Bigr)               \, ,
\eeq
where $y$ are the complex coordinates describing the D3-brane position in the internal space, and $k(y,\bar{y})$ is the K\"ahler potential for the Calabi-Yau metric on this space.  For the case of D3-branes moving on the conifold, the explicit form of $k$ is known \cite{Candelas:1989js}, but will not be needed here.

Elementary counting of equations shows that in generic situations, the F-terms for the $\rho,y$ system vanish only at special points in the configuration space, {\it i.e.}~solutions to $D_\rho W = D_y W=0$ are isolated.
For configurations involving moduli-stabilizing D7-branes in the throat region, specific examples of isolated vacua were found in \cite{DeWolfe}.  Our present interest is in estimating the scale of the F-term potential experienced by a D3-brane in the UV region of a throat that is glued into a nonperturbatively-stabilized bulk,
in the case that all divisors responsible for moduli stabilization are outside of the throat.

To this end, we further note that the nonperturbative superpotential vanishes whenever a mobile D3-brane sits on the divisor bearing the nonperturbative effects \cite{Ganor}.  (In the presence of positive-energy sources, such as anti-D3-branes, there is consequently a generic decompactification instability.)  On the other hand, when the D3-brane sits in one of its isolated supersymmetric vacua, the four-dimensional cosmological constant is negative.\footnote{For this discussion we omit the special case of supersymmetric Minkowski vacua.} It follows that the change of four-dimensional potential energy required to move a D3-brane from a supersymmetric vacuum location, onto the divisor responsible for K\"ahler moduli stabilization, is of order the cosmological constant $V_{AdS}$ at the minimum of the potential,
\beq \label{equ:change}
        \Delta V_F(D3) \sim    |V_{AdS}|   \, .
\eeq

We next consider other sources of supersymmetry breaking.  In addition to the contribution of the brane-antibrane pair responsible for the inflationary energy, there must in general be an additional `uplifting' contribution ({\it e.g.}~from a distant source of supersymmetry breaking, such as other anti-D3-branes) in order for the cosmological constant to remain positive after the annihilation of the inflationary brane-antibrane pair.  A model-building assumption that always underlies this scenario is that the net positive energy $V_{uplift}$ introduced by all such sources of supersymmetry breaking, including the inflationary brane-antibrane pair, is not so large that it causes decompactification.  From the shape of the barrier to decompactification, one finds \cite{KKLT}
  $      V_{uplift} \lesssim {\rm few}\times |V_{AdS}|  $.
This condition can be relaxed in more complicated configurations,
{\it e.g.}~\cite{KL}.  The inflationary energy can be an ${\cal{O}}(1)$ fraction of $V_{uplift}$, or can be considerably smaller, so that we may write
\beq \label{equ:ratio}
         V_{D3/\overline{D3}}= 2 a_0^4 T_3 = \varepsilon |V_{AdS}|   \, , \qquad \varepsilon \lesssim {\cal O}(1)\, .
\eeq

We therefore conclude that, in  the metastable de Sitter vacua of \cite{KKLT}, the typical change in the F-term potential upon moving a D3-brane from its supersymmetric minimum to the divisor location is
\beq \label{equ:scaling}
            \Delta V_F(D3) \sim | V_{AdS}| \sim \frac{1}{\varepsilon}\times 2 a_0^4 T_3  \, .
\eeq

Once again, our goal is to determine the typical scale $V_{\rm UV}(D3)$ of the F-term potential for a D3-brane in the UV region of the throat. In general, $V_{\rm UV}(D3) \lesssim \Delta V_F(D3)$. The approximate equality applies when the divisor bearing nonperturbative effects enters the throat region; but in this case, the analysis reduces to that of \cite{delicate}.  On the other hand, if the divisor is well-separated from the throat,  and instead exists exclusively in the bulk region,  a D3-brane inside the throat will feel an F-term potential that is suppressed compared to the estimate (\ref{equ:scaling}): the interaction with the divisor is of course suppressed at large distances, as can be seen {\it e.g.}~from the increasing mass of strings stretching from the D3-brane to the divisor.

We conclude that the scale of the D3-brane F-term potential is variable and depends on the proximity of the moduli-stabilizing divisor to the throat region.  This is conveniently represented by
\beq \label{equ:ans}
                   V_{\rm UV}(D3) = \frac{c_F}{\varepsilon} 2 a_0^4 T_3 \equiv  c\, a_0^4 T_3    \, ,
\eeq
where the coefficient $c_F$ can be small if the divisor is far from the throat, and is in any case bounded above by $c_F \lesssim 1$, following (\ref{equ:scaling}).  Clearly, by adjusting the structure of the compactification and the scales of the various sources of supersymmetry breaking, one can arrange to vary $c$ in a wide range.

%%%%%%%%%%%%%%%%%%%%%%%%%%%%%%%%%%%%%%%%%%%%%%%%

\end{document}